# Spin-pumping into surface states of topological insulator α-Sn, spin to charge conversion at room temperature

J.-C. Rojas-Sánchez[1,2]*,   S. Oyarzun[3,4],   Y. Fu[3,4],   A. Marty[3,4],   C. Vergnaud[3,4],   S. Gambarelli[3,4],   L. Vila[3,4], M. Jamet[3,4], Y. Ohtsubo[5,6],   A. Taleb-Ibrahimi[7,8],   P. Le Fèvre[8], F. Bertran[8], N. Reyren[1,2]*, J.-M. George[1,2]  and A. Fert[1,2]

1 Unité Mixte de Physique CNRS/Thales, 91767 Palaiseau, France

2 Université Paris-Sud, Université Paris-Saclay, UMR137, 91767 Palaiseau, France

3 Université Grenoble Alpes, INAC-SP2M, F-38000 Grenoble, France

4 CEA, Institut Nanosciences et Cryogénie, F-38000 Grenoble, France

5 Graduate School of Frontier Biosciences, Osaka University, Suita 565-0871, Japan

6 Graduate School of Science, Osaka University, Toyonaka 560-0043, Japan

7 UR1 CNRS, Synchrotron SOLEIL, Saint-Aubin, 91192 Gif sur Yvette, France

8 Synchrotron SOLEIL, Saint-Aubin, 91192 Gif sur Yvette, France

* contact email:   juan-carlos.rojassanchez@thalesgroup.com,            nicolas.reyren@thalesgroup.com

We present experimental results on the conversion of a spin current into a charge current by spin pumping into the Dirac cone with helical spin polarization of the elemental topological insulator (TI) α-Sn[1-3]. By angle-resolved photoelectron spectroscopy (ARPES) we first confirm that the Dirac cone at the surface of α-Sn (0 0 1) layers subsists after covering with Ag. Then we show that resonant spin pumping at room temperature from Fe through Ag into α-Sn layers induces a lateral charge current that can be ascribed to the Inverse Edelstein Effect[4-5]. Our observation of an Inverse Edelstein Effect length[5-6] much longer than for Rashba interfaces[5-10] demonstrates the potential of the TI for conversion between spin and charge in spintronic devices. By comparing our results with data on the relaxation time of TI free surface states from time-resolved ARPES, we can anticipate the ultimate potential of TI for spin to charge conversion and the conditions to reach it.

The two-dimensional electronic states at the interfaces with Rashba interactions (as, for example, Bi|Ag[11]) or at the surface/interface of the materials called topological insulators[12-14] are characterized by Fermi contours with helical locking of spin with momentum by spin-orbit interactions. Figure 1a-b presents a schema of the dispersion surfaces of Rashba states (a) and Dirac cones (DC) of TI surface/interface states (b), as well as the helical spin configurations of the Fermi contours. As depicted in Fig.1c-d, a current carried by such two dimensional electron gas (2DEG) with helical spin polarization is automatically accompanied by a nonzero spin accumulation along the in-plane direction transverse to the current. This effect has been predicted in 1990 by Edelstein[4] and can be described as a charge to spin conversion effect. The Edelstein Effect (EE) has been recently demonstrated by spin-polarized positron experiments[15] for the Rashba interface Bi|Ag and is also involved in spin torque experiments with TI[16].

The Inverse Edelstein Effect[5,6,17] (IEE) can be described as the inverse conversion of the one in EE. As depicted in Fig.1e-f, the injection of a vertical spin current into the 2DEG at a Rashba or TI surface/interface induces a charge current in the 2DEG. The IEE length[5] $\lambda_{\text{IEE}}$ is the ratio between the 2D conventional charge current density

$j_C^{2D}$ (in A/m) induced by IEE in the surface/interface 2DEG and the injected 3D spin current density, $j_{S(i)}^{3D}$. We

adopt the usual definition with the injected spin current density with $j_{S(i)}^{3D}$ equal to the difference between the injected charge current densities carried by electrons having their spin respectively oriented along the +$i$ and −$i$ directions along the $x$- or $y$-axis (the corresponding injected spin flow density is $j_{S(i)}^{3D}/(2e)$ where $e = −|e|$). For both Rashba and TI interfaces, and in the simple situation of circular spin contours, $\lambda_{\text{IEE}}$ can be expressed as a function of the relaxation time $\tau$ of an out of equilibrium distribution in the topological states by the following expressions: $\lambda_{\text{IEE}} = \alpha_R \tau / \hbar$ for Rashba interfaces[5,6], where $\alpha_R$ is the Rashba coefficient, and, as derived in Supplementary Information (SI1),

$$\lambda_{\text{IEE}} = v_F \tau \tag{1}$$

for TI, where $v_F$ is the Fermi velocity of the DC. To be more precise on the sign, our definition of the IEE length is exactly

$$\lambda_{\text{IEE}} = j_C^{2D(x)} / j_{S(y)}^{3D} = -j_C^{2D(y)} / j_{S(x)}^{3D} \ , \tag{2}$$

where the upper (lower) index refers to the current direction (spin quantization axis). With this definition a positive $\lambda_{\text{IEE}}$ corresponds to an Inverse Spin Hall effect (ISHE) with a positive spin Hall angle (SHA), as the one of Pt. The IEE length of TI is expected to be positive for the situation of spin injection into a counterclockwise helical Fermi contour of electrons (Fig.1a-b).

The spin to charge conversion (SCC) by IEE was demonstrated by experiments of resonant spin pumping[18] onto interfaces, first with Rashba interfaces[5-10] and then with TI by Shiomi *et al*[17]. Here we present similar spin pumping experiments on α-Sn (0 0 1) films that reveal very efficient SCC by IEE at the interface of this TI. The SCC by the EE and IEE is promising for the generation and detection of spin currents in spintronic devices, especially if they can be obtained with TI interfaces that, compared to Rashba interfaces, do not suffer from the partial compensation of the two Fermi contours of opposite helicity. For applications, an efficient conversion subsisting at room temperature is also mandatory.

Band gap opening and TI properties can be induced in α-Sn (0 0 1) films either by strain[1-2] or quantum-size effects in thin films.[1,3] The recent ARPES measurements of Ohtsubo *et al.*[3] were performed on the Cassiopee beam line at Synchrotron SOLEIL on thin α-Sn (0 0 1) films grown *in-situ* by molecular beam epitaxy on treated InSb (0 0 1) substrate. They revealed a Dirac Cone (DC) linear dispersion with helical spin polarization around the $\Gamma$- point of the surface Brillouin zone. The Fermi energy $E_F$ is in the lower cone at 20 meV below the Dirac Point (DP) for 24 monolayers (ML) and in the upper cone at around 50-100 meV above DP for 30-34 ML. The spin-resolved ARPES indicated the clockwise (CW)/counterclockwise (CCW) spin helicity in the lower/upper cone, respectively. All our InSb(001)|α-Sn(001)|Fe and InSb(001)|α-Sn(001)|Ag|Fe samples were grown in the same conditions (see Methods) at the Cassiopee beamline in order to check by ARPES if the DC subsists after covering by Fe or Ag and before taking off the samples for spin pumping.

In the ARPES images of Fig. 2, a DC is clearly seen at the free surface (top) of our α-Sn(001) samples (around 30 ML), it disappears for coatings of Fe as thin as half of a monolayer, *i.e.* 0.09 nm (left), but subsists when α-Sn is covered with even 1.2 nm of Ag (right). Without and with Ag, $E_F$ is in the upper cone as in the results of Ohtsubo et al[3] for 30-34 ML. The quantitative analysis of the ARPES plots with Ag reveals that ($E_F$−$E_{DC}$) is in the range 50-85 meV and $v_F$ around $6\times10^5$ m/s (Supplementary Fig. 2). We can thus expect that only the α-Sn/Ag/Fe samples will show SCC by IEE. This is confirmed by the results displayed in Fig. 3b-c: *i*) A large enhancement of the damping coefficient revealing significant spin absorption is seen in Fig. 3b only for α-Sn/Ag/Fe and not for α-Sn/Fe. *ii*) In Fig. 3c, a dc charge current $I_C$ peak at the resonance is only seen for α-Sn/Ag/Fe.

We used the standard analysis[5,18] of spin pumping (Supplementary Discussion 3) to extract the injected spin current density $j_S^{3D}$ and finally the IEE length $\lambda_{IEE}$=2.1 nm from equation (2) (with $j_C^{2D} = I_C$ /(α-Sn width)). The positive sign of $\lambda_{IEE}$ is consistent (Supplementary Discussion 1) with the CCW helical spin configuration of the electron-type Fermi contours in[3] α-Sn. This value of $\lambda_{IEE}$ is one order of magnitude larger than $\lambda_{IEE}$≈0.2-0.4 nm found at the Bi/Ag Rashba interface. It also shows that the conversion by the α-Sn TI is much more efficient than what can be expected by ISHE. The conversion by ISHE, characterized by the SHA $\Theta_{SHE}$, *i.e.* the ratio between 3D charge and 3D spin current densities, can be compared to the conversion by IEE if we integrate the ISHE-induced 3D charge current density from top to bottom of the SHE layer to obtain the 2D charge current density. The maximum value for $t>>l_{sf}$ is $j_C^{2D}$(ISHE) $= \Theta_{SHE} l_{sf} j_S^{3D}$ and corresponds to an IEE-like length $\lambda^* = \Theta_{SHE} l_{sf}$. For Pt, with $l_{sf}$≈3.4 nm[19] and $\Theta_{SHE}$≈0.056[19], $\lambda^*$≈0.19 nm and for thin W films[20-21] with[21] $l_{sf}$≈1.4 nm and[20] $\Theta_{SHE}$≈0.33 or[21] 0.19, $\lambda^*$≈0.46 or 0.27 nm. In other words the value 2.1 nm of α-Sn would correspond to $\Theta_{SHE}$≈0.62 for Pt instead of 0.056 and 1.5 instead of 0.33 or 0.19 for W.

An important parameter in equation (1) is the relaxation time $\tau$ of out-of-equilibrium distributions in the topological states. From $\lambda_{IEE}$≈2.1 nm and $v_F$≈$6\times10^5$ m/s we find $\tau$≈3.5 fs. Interestingly the values of $\tau$ derived from spin pumping experiments on TI or Rashba "interfaces" (5 fs for the Bi/Ag Rashba interface[5]) are definitely shorter than relaxation times in the picosecond range derived from time-resolved ARPES measurements on "free surfaces" of TI[22]. We believe that the long relaxation time on free topological surfaces characterizes the slow relaxation inside the 2D topological states whereas, by interfacing the TI (or Rashba) surface with a metal as Ag, we introduce a faster additional relaxation mechanism provided by exchanges of electrons with the adjacent 3D metal. We conclude that the best conditions for the exploitation of topological sates of TI in spintronics should be with TI interfaced with (trivial) insulators instead of metals, for example in experiments of spin-pumping or thermal[23] spin injection from an insulating ferromagnet. If one remarks that $v_F \tau$ is also the critical length for ballistic transport, one can also anticipate that ballistic transport should be generally limited to the nanometer range at a TI/metal interface but can probably reach the micrometer range on free surfaces and possibly at interfaces with insulating materials.

In summary, a very efficient spin to charge current conversion (SCCC) is achieved by spin-pumping into topological insulator α-Sn films, in clear relation with the IEE induced by the CW helical spin configuration of the Dirac cones that we have identified by ARPES at the α-Sn (0 0 1)Ag interface. To our knowledge, these results are the first of IEE and SCCC by a TI at room temperature. They open the road not only to deeper characterizations of IEE with α-Sn (dependence on thickness, temperature and gate voltage, not in the scope of this letter) but also to the exploitation of the IEE in spintronic devices at room temperature. The final result, the much shorter relaxation time of the topological states at the α-Sn (0 0 1)|Ag interface in comparison with the relaxation time at TI free surfaces derived from time-resolved ARPES is probably due to the electron exchange with the adjacent 3D metal layer, allowing us to anticipate much more efficient IEE at the interface of TI with insulators.

## Methods

**Samples fabrication.** The films were grown by Molecular Beam Epitaxy (MBE) on crystalline (0 0 1)-oriented InSb with a $c(8\times2)$ reconstructed surface. The deposition rate of α-Sn, Fe, Ag and Au were around 1.2 Å/min (0.27 ML/min), 0.13 Å/s, 1.6 Å/min, and 3.4 Å/min, respectively. The thickness of α-Sn was monitored *in-situ* by Reflection High-energy Electron Diffraction (RHEED) and verified *a posteriori* in some of the samples by electron microscopy analysis (no shown). Details on the InSb surface preparation and α-Sn growth are described in ref. 3.

**ARPES measurements.** The ARPES measurements were performed at room temperature with incident photon energy of 19 eV and resolving angle between ±15° which correspond to wave number $k$ between ±5 nm$^{-1}$ at the Fermi level. In Fig. 2, only the area of interest is shown.

**Ferromagnetic resonance (FMR) and spin pumping**. The samples have the stacking order shown in Fig. 3. The broadband frequency dependence was performed in a coplanar wave guide system, applying the external magnetic film at different in-plane crystalline directions of the substrate. The samples were then cut in slab of 2.4x0.4 mm to carry out the simultaneously FMR and transversal dc voltage measurement (Fig. 3a,c). The slab is placed on the axis of a cylindrical X-band cavity (frequency ≈ 9.6 GHz). The charge current $I_C$ is derived from the voltage $V$ needed to cancel it, $I_c = V/R$ where $R$ is the resistance of the sample measured between the voltage probes.

# Acknowledgements

We acknowledge Laura Bocher, Katia Marsch and Odile Stéphan from the "Laboratoire de Physique des Solides" (Université Paris-Sud) and David Troadec from the "Institut d'Electronique, de Microélectronique & de Nanotechnologie" (CNRS) for their help during the characterization of the samples. We thank I. Pheng for the preparation of the sample for SP-FMR.

## Author Contributions

JCRS, JMG, ATI and AF proposed the study and supervised the project. JCRS, YO, NR, PLF, FB, and JMG, grown the samples and acquired the *in situ* ARPES data with the help of ATI, CV, AM, and MJ. The ARPES data was analyzed by NR and JCRS with the help of YO, SO, PLF and FB. The spin-pumping experiment was performed by JCRS, SO, YF, SG, LV and JMG. JCRS, NR, YO, JMG and AF analyzed all the data and wrote the letter. All the authors comment on the manuscript.

## Competing Financial Interests

The authors declare no competing financial interests.

## Figure Legends

**Figure 1 | Edelstein and Inverse Edelstein Effect**. **a-b**, Top: Energy dispersion surfaces of the 2D states at a Rashba interface (a) and Dirac dispersion cone of the surface/interface states of a topological insulator (b). Bottom: Fermi contours of Rashba states (a) with two contours of opposite chirality of their helical spin configurations and TI surface/interface states (b). **c-d**, Edelstein Effect: A flow of electrons along $x$ ($j_C^{2D(x)} < 0$ in the figure) in Rashba (c) or TI (d) 2DEGs is associated with shifts $\Delta k$ of the Fermi contours generating an extra-population of spin along the $y$ direction (for Rashba there is a partial compensation of the opposite spin directions generated by the two contours). **e-f**, Inverse Edelstein Effect: Injection of a spin current density spin-polarized along $y$ ($j_{S(y)}^{2D} < 0$ in the figure) into Rashba (e) or TI (f) 2DEGs induces an extra-population on one side of the Fermi contour (along the x direction), a depletion on the other side and therefore a charge current density $j_C^{2D(x)}$. $j_{S(y)}^{2D}$ and $j_C^{2D(x)}$ of this figure can be seen as carried by the electrons of wave vector (wavy arrow) and spin (straight arrow) in the schematic at bottom of (a-b). The expressions of the IEE length $\lambda_{IEE}$ relating the 2D charge current to the 3D injected spin current are given in the bottom-right of the figure as a function of the relaxation time of the topological states and Rashba coefficient $\alpha_R$ or Fermi velocity $v_F$ of the TI Dirac cone. All the figures are drawn for electron-type conduction in the 2D states.

**Figure 2 | ARPES characterization of the interfaces states of α-Sn (0 0 1) films covered by Fe or Ag**. Top: ARPES plot of the DC along [100] or [110] on the free surface of α-Sn (approximately 30 ML thick). Below:

ARPES plots after covering by 0.9 Å or 1.8 Å of Fe (left) and 4.3 Å or 12 Å of Ag (right). The intensity is color-coded with the same color-scale in each panel (arbitrary units).The DC subsists if α-Sn is covered by Ag and it can still be seen when the Sn|Ag interface is buried below 12 Å of Ag. It disappears after covering by Fe but the overlap with the high intensity of the Fe 3$d$ band dispersion at the Fermi level precludes definitive conclusions on the complete suppression of the DC.

**Figure 3 | Spin-to-charge current conversion by spin pumping onto the topological states of α-Sn (0 0 1). a**, Experimental setup for spin pumping into α-Sn by Ferromagnetic Resonance (FMR) of a Fe layer (see Methods). **b**, Broadband frequency dependence of the peak-to-peak FMR linewidth when the magnetic field $H$ is applied along the [1 0 0] direction of the Insb substrate for InSb|Fe|Au (reference for the Fe layer), InSb|αSn|Fe|Au and InSb|αSn|Ag|Fe|Au samples. The effective damping coefficient of Fe, proportional to the slopes, is definitely larger with Ag interface (α=0.028 compared to 0.008 and 0.006 for the two other structures) which show the strong spin absorption by the Ag|Sn interface. **c**, FMR and dc charge current signals from measurements in a cylindrical X-band resonant cavity on InSb|Fe|Au, InSb|α-Sn|Fe|Au and InSb|α-Sn|Ag|Fe|Au samples. Only the third sample shows a dc current signal in agreement with the observation of a Dirac cone at only the Sn|Ag interface and the resulting spin to charge conversion by IEE.

**a** Rashba Interface

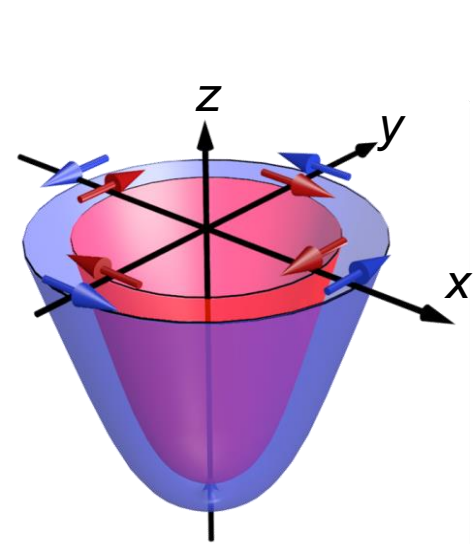

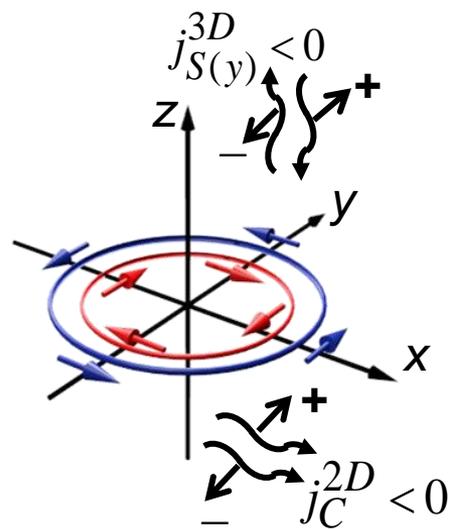

$$j_{S(y)}^{3D} < 0$$

$$j_C^{2D} < 0$$

**b** TI Interface

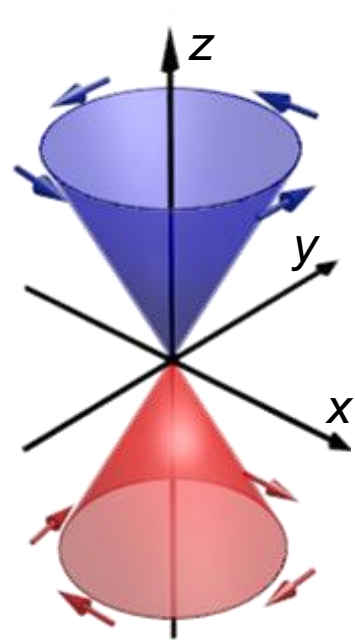

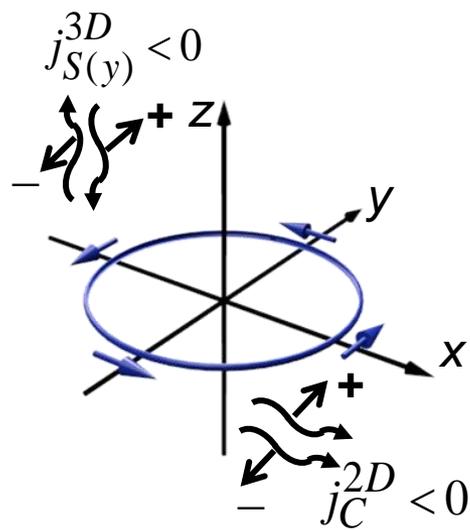

$$j_{S(y)}^{3D} < 0$$

$$j_C^{2D} < 0$$

**c** Edelstein Effect (EE) **d**

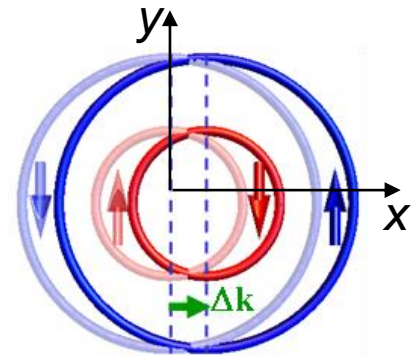

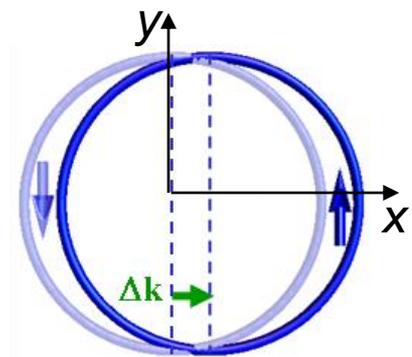

$\Delta k$

$\Delta k$

**e** Inverse Edelstein Effect (IEE) **f**

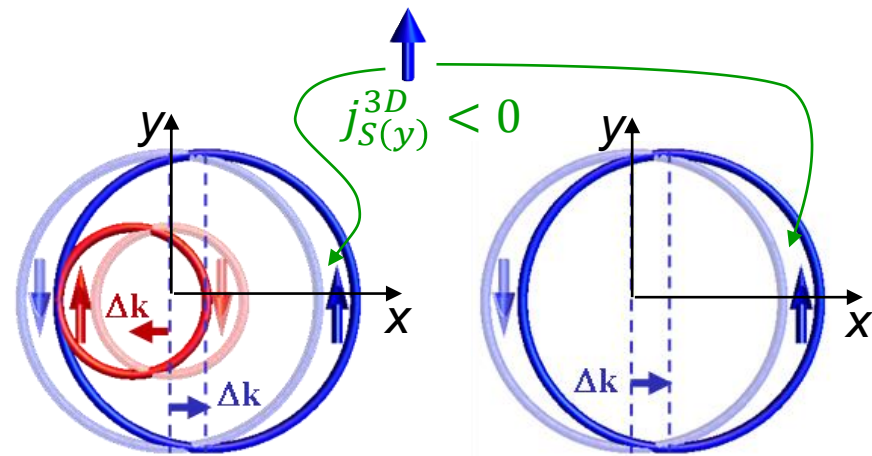

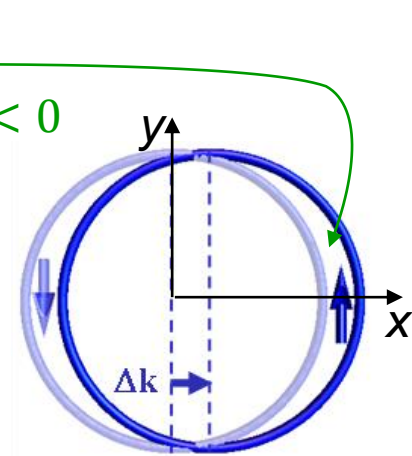

$$j_{S(y)}^{3D} < 0$$

$\Delta k$

$\Delta k$

$$j_C^{2D} = \lambda_{IEE}\, j_S^{3D}$$

$$\lambda_{IEE} = \frac{\alpha_R \tau}{\hbar}$$

$$j_C^{2D} = \lambda_{IEE}\, j_S^{3D}$$

$$\lambda_{IEE} = v_F \tau$$

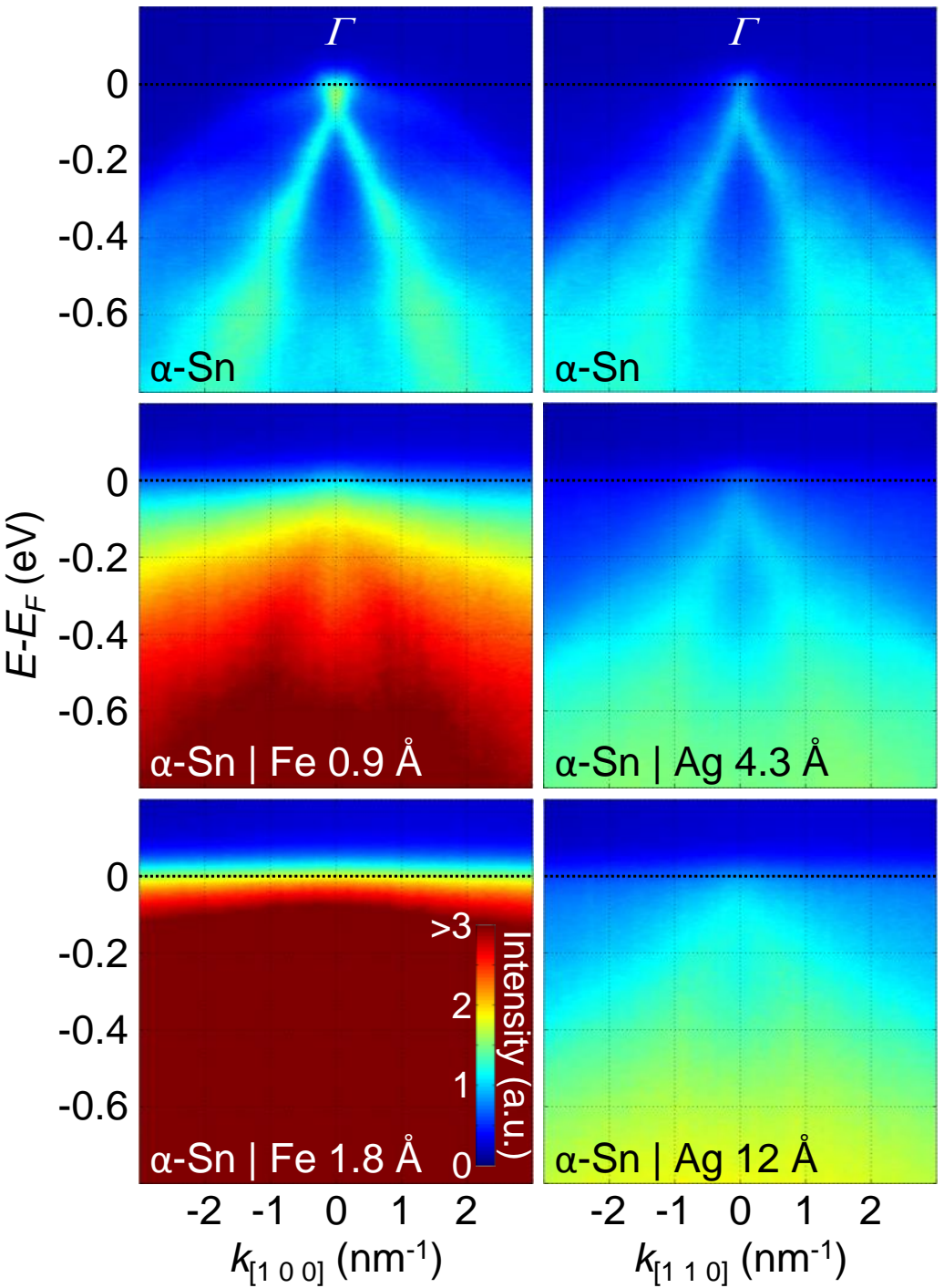

**a**

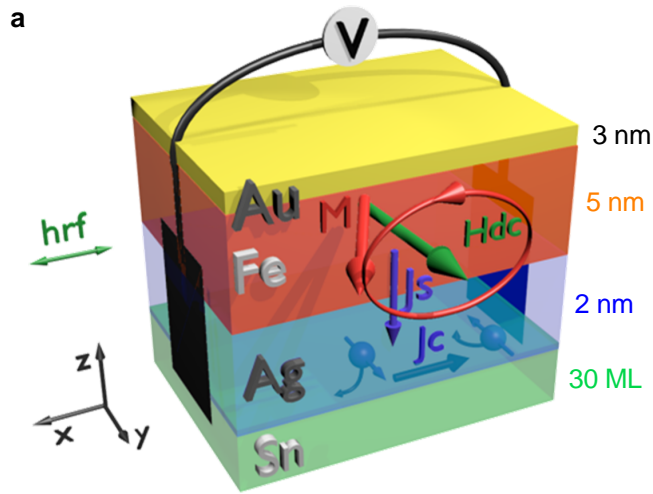

**b**

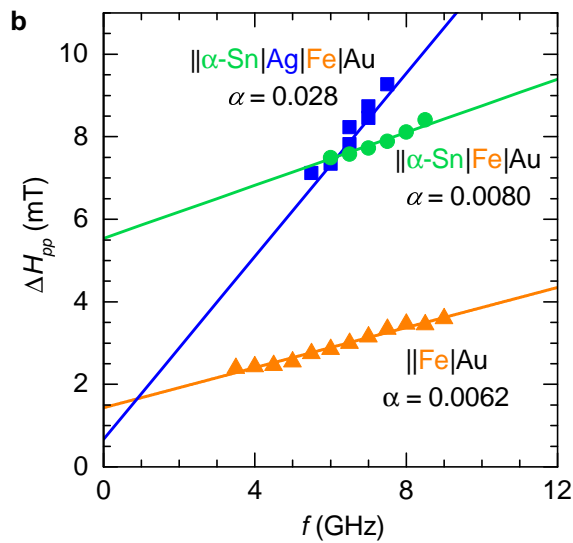

**c**

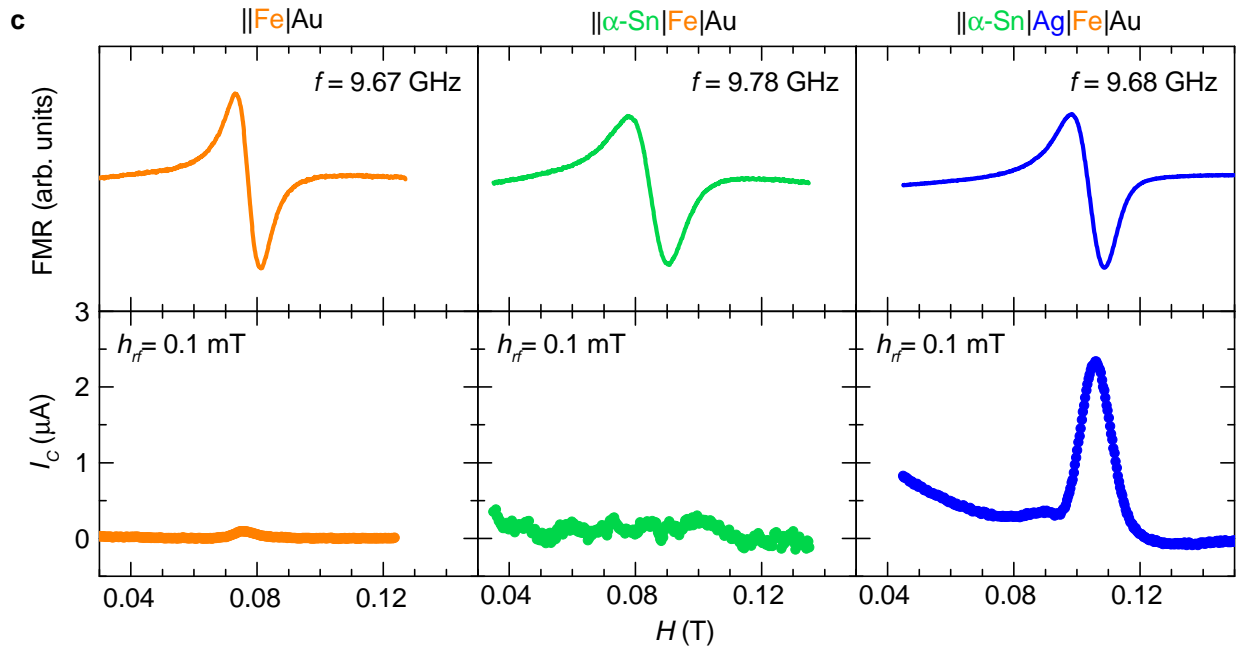

# Spin-pumping into surface states of topological insulator α-Sn, spin to charge conversion at room temperature

J.-C. Rojas-Sánchez[1,2], S. Oyarzun[3,4], Y. Fu[3,4], A. Marty[3,4], C. Vergnaud[3,4],

S. Gambarelli[3,4], L. Vila[3,4], M. Jamet[3,4], Y. Ohtsubo[5,6], A. Taleb-Ibrahimi[7,8], P. Le Fèvre[8],

F. Bertran[8], N. Reyren[1,2], J.-M. George[1,2] and A. Fert[1,2]

[1] Unité Mixte de Physique CNRS/Thales , 91767 Palaiseau, France

[2] Université Paris-Sud, Université Paris-Saclay, UMR137, 91767 Palaiseau, France

[3] Université Grenoble Alpes, INAC-SP2M, F-38000 Grenoble, France

[4] CEA, Institut Nanosciences et Cryogénie, F-38000 Grenoble, France

[5] Graduate School of Frontier Biosciences, Osaka University, Suita 565-0871, Japan

[6] Graduate School of Science, Osaka University, Toyonaka 560-0043, Japan

[7] UR1 CNRS, Synchrotron SOLEIL, Saint-Aubin, 91192 Gif sur Yvette, France

[8] Synchrotron SOLEIL, Saint-Aubin, 91192 Gif sur Yvette, France

**Supplementary Discussion 1 | Inverse Edelstein Effect length for Topological Insulators (TI)**

We suppose, as in Fig. 1b (bottom), a spin current density $j_{S(y)}^{3D}$ pumped into the 2DEG at the TI interface, that is, with the definition of $j_{S(y)}^{3D}$ introduced in the text, an incoming flow of spins oriented along $y$ of density $j_{S(y)}^{3D}/(2e)$ (with $e = -|e|$). With the CCW spin configuration of the electron-like and circular electron Fermi contour of the figure, the injection (extraction) of electrons of spin along $y$ ($-y$) on empty (from occupied) states on the $x$ ($-x$) side of the Fermi contour induces an out-of equilibrium distribution (shift $\Delta k$ of the Fermi contour in first approximation) with a nonzero spin density in the $y$ direction and a charge current in the $-x$ direction. We call $\tau$ the relaxation time of this out-of-equilibrium distribution in the topological states that associates a non-zero spin density $<\delta\sigma_y>$ and a shift of the Fermi contour in the momentum space.

In the general case, for both CW and CCW helicities and also for both electron- and hole-like Fermi contours, the balance between the incoming spin flow and the relaxation rate can be written as

$$j_{S(y)}^{3D}/2e = <\delta\sigma_y>/\tau \quad. \tag{S1}$$

In the simplest situation with circular Fermi contour shifted by $\Delta k \ll k_F$, the spin density $<\delta\sigma_y>$ and the 2D charge current density $j_C^{2D(x)}$ associated with the same $\Delta k$ can be calculated straightforwardly and we obtain for their ratio

$$j_C^{2D(x)}/<\delta\sigma_y> = \pm 2e v_F \tag{S2}$$

where $v_F$ is the Fermi velocity of the Dirac cone. In the situation with an electron type Fermi contour (upper cone), + is for CCW and − for CW for electron-like Fermi contour and the opposite signs for hole-like Fermi contour (we remind: $e = -|e|$). Finally, by combining equations (S1) and (S2), we obtain the IEE length:

$$\lambda_{IEE} \equiv j_C^{2D(x)}/j_{S(y)}^{3D} = \pm v_F \tau \quad, \tag{S3}$$

where + is for electron-like Fermi contour (FC) with CCW helicity or hole-like FC with CW helicity and − for the two other combinations. The experimental situation of our spin pumping experiments on α-Sn correspond the electron-like FC (upper cone) with CCW helicity, with a positive $\lambda_{IEE}$ in agreement with equation (S3). The helicity is defined from the point of view of an observer on the injection side.

We finally want to point out that the above results are not strictly original as they are equivalent to results expressed differently by Shiomi et al[17] or for different situations by Culcer[S1]. Another final remark about the IEE being that the injection of a spin current polarized along y not only induces a charge current along x but also generates, in some conditions, a spin current along y spin-polarized in the direction $z^{6,S1}$, an additional SHE-like effect not in the scope of the letter.

## Supplementary Discussion 2 | Dirac Cone (DC) and Fermi velocity

The Fermi velocity of the Dirac cones of our α-Sn-based samples is derived as it is shown in Figure S1. This velocity estimated for the 30 ML sample with the lowest Ag thickness in the figure is around $6\cdot10^5$ m/s which is 20% smaller than the value reported in the work of Ohtsubo et al[3] for the free surface of a 24 ML α-Sn film.

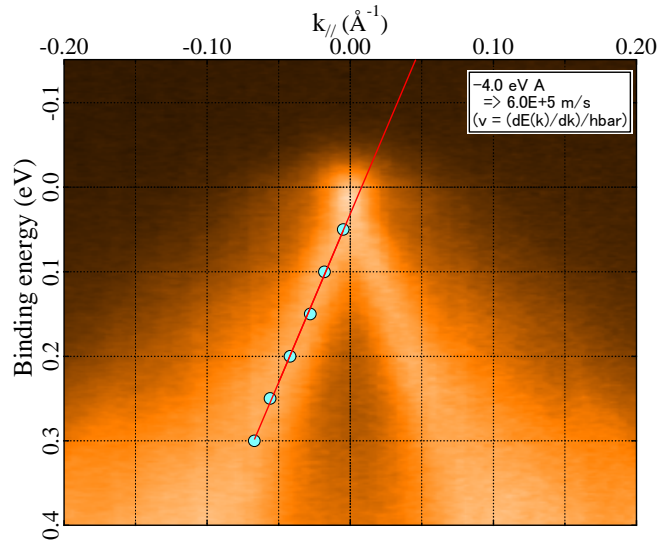

**Figure S1 |Dirac cone and Fermi velocity from ARPES.** The dots correspond to the highest intensity in horizontal scans and the Fermi velocity of the Dirac cone is derived from the slope of the line between the dots.

## Supplementary Discussion 3 | Analysis of the FMR spin pumping experiments

In Figure S2a we show the results of the broadband field dependence of the resonance frequency for three different systems when the magnetic field *H* is applied along the [1 0 0] crystalline orientation of the substrate. The field dependence of the frequency is typical of a FM thin film with cubic and in-plane uniaxial anisotropies and the difference between the α-Sn|Ag|Fe samples and the two others can be explained by the different anisotropy constant for Fe grown on Ag.

For the derivation of the injected spin current density we refer to the classical analyses of spin pumping (by Ando et al[18] for example) or to the very similar spin pumping and IEE experiments of some of us[5] with Bi|Ag Rashba interfaces. From the results of these previous experiments on Bi|Ag, we also know that we can neglect the spin absorption by the thin Ag layers between the ferromagnetic layer and the IEE active α-Sn interface. Following such a formalism, we have estimated the effective spin mixing conductance to be $g_{eff}^{\uparrow\downarrow} = 70.7$ nm$^{-2}$, and the spin current density injected at the resonance condition is $j_S^{3D} = 2.8$ MA/m$^2$ if the strength of the rf field is $h_{rf}$ = 0.1 mT. From the amplitude of the symmetrical Lorentzian voltage, the current is $I_C$= 2.37 μA, and $J_C^{2D = 5.9}$ mA/m (values already normalized for a $h_{rf}$=0.1 mT). Then it is quite straightforward to calculate $\lambda_{IEE}$=2.1 nm following equation (2) in the main text. The experimental sign of $\lambda_{IEE}$>0 have been also verified by directly comparison with the measurements in the same setup of SP-ISHE in different materials as in Pt[18].

An important point to clarify is the possible contributions from Seebeck and spin-Seebeck effects[23]. The dc voltage signal generated by the Seebeck effect in the presence of a FMR-induced horizontal temperature gradient is expected to have the same sign for opposite magnetic fields, in contrast with the IEE signal that is reversed when the field is reversed. To suppress any possible Seebeck contribution we have always averaged the voltage signals obtained at +H and −H, <V>=($V_{H+}$−$V_{H-}$)/2. The possible contribution to the voltage signal from a FMR-induced vertical temperature gradient inducing an additional spin current and the resulting Spin Seebeck voltage has opposite signs for +H and −H but, as clearly shown in the spin pumping experiments of Shiomi *et al*[17] on TI, its dependence on the rf power departs markedly from the linear variation of the IEE signal. We can rule out any significant spin-Seebeck contribution from the linear variation of the voltage signal with the rf power depicted in Figure S2 for one of our samples.

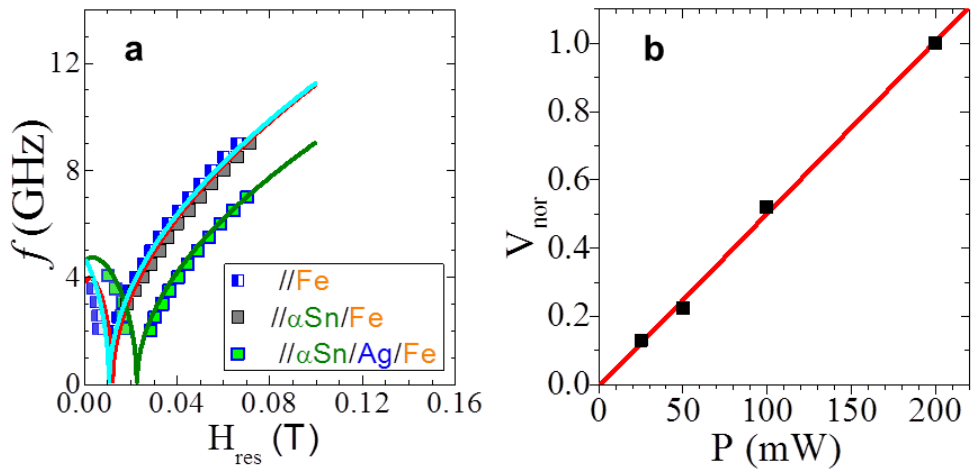

**Figure S2 | Ferromagnetic resonance and spin pumping data. a,**Broadband frequency dependence of resonance field in FMR measurements carried out in a coplanar waveguide. The lines correspond to numerical simulations assuming, in addition to Zeeman and shape magnetic anisotropy, uniaxial in-plane and cubic magnetocrystalline anisotropies. **b,** Power dependence of the voltage amplitude measured at resonance condition in the X-band cylindrical cavity for one of the α-Sn|Ag|Fe samples. $V_{nor}$ = $V/V$(200mW).